# Detecting LLM-assisted writing in scientific communication: Are we there yet?

Teddy Lazebnik[1,2], Ariel Rosenfeld[3†]

[1] Department of Mathematics, Ariel University, Ariel, Israel
[2] Department of Cancer Biology, Cancer Institute, University College London, London, UK
[3] Department of Information Science, Bar Ilan University, Ramat Gan, Israel

**Abstract**

Large Language Models (LLMs), exemplified by ChatGPT, have significantly reshaped text generation, particularly in the realm of writing assistance. While ethical considerations underscore the importance of transparently acknowledging LLM use, especially in scientific communication, genuine acknowledgment remains infrequent. A potential avenue to encourage accurate acknowledging of LLM-assisted writing involves employing automated detectors. Our evaluation of four cutting-edge LLM-generated text detectors reveals their suboptimal performance compared to a simple ad-hoc detector designed to identify abrupt writing style changes around the time of LLM proliferation. We contend that the development of specialized detectors exclusively dedicated to LLM-assisted writing detection is necessary. Such detectors could play a crucial role in fostering more authentic recognition of LLM involvement in scientific communication, addressing the current challenges in acknowledgment practices.

## 1 Introduction

Sophisticated Large Language Models (LLMs), such as ChatGPT, have become highly effective in comprehending and generating human-like texts and have become pivotal in various applications, including writing assistance (Seßler et al., 2023). From an ethical standpoint, appropriately acknowledging the use of LLMs is conscientious as it underscores a commitment to transparency, honesty, and integrity in writing (Sallam, 2023). In scientific communication, where the pursuitand dissemination of knowledge are deeply guided by these principles (Sikes, 2009), clearly articulating the involvement ofthese models in the writing process is of great importance (Bin-Nashwan et al., 2023). Indeed, such acknowledgment is mandated by mostpublishers (Nature editorial, 2023).

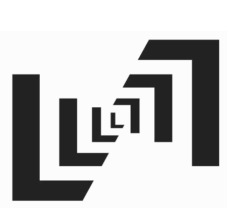

† Corresponding author: Ariel Rosenfeld ( Email: ariel.rosenfeld@biu.ac.il).

Unfortunately, the current landscape presents a formidable challenge in terms of enforcing the explicit acknowledgment of LLMs in general and in scientific communication in particular. First, the evolving capabilities of LLMs and their intricate role in the writing process introduce uncertainty and ambiguity, thus estimating the extent of LLM influence and establishing clear boundaries for acknowledgment can be elusive. In addition, certain authors may hesitate to overtly admit their use of LLM for various reasons, including a traditional authorship viewpoint, concerns about potential negative perceptions, or the lack of established guidelines in this regard, to name a few (Draxler et al., 2023).

One promising strategy to promote genuine disclosure of LLM usage involves the development of automated tools for LLM use detection. Indeed, various detectors were created for distinguishing between human and LLM-generated texts (Tang et al., 2023). Nevertheless, these detectors need not necessarily be proficient in detecting LLM-assisted writing, as they were not originally designed for this purpose. To our knowledge, the automated detection of LLM-assisted writing in scientific communication has yet to be explicitly considered by the scientific community.

In this work, we investigate the viability of using four state-of-the-art LLM-generated text detectors for detecting LLM-assisted writing in scientific communication. Our findings reveal subpar performance, raising substantial concerns regarding the practical value of using these models to detect potentially undisclosed LLM-assisted writing. To make the case for the viability of the LLM-assisted writing detection challenge, we present and evaluate an alternative detector designed for identifying abrupt "writing style changes" occurring around the period of LLM proliferation, which can reasonably be associated with LLM writing. While our proposed approach need not claim optimality and is limited in several respects, it does demonstrate a noteworthy improvement compared to existingdetectors, indicating that the challenge of detecting LLM-assisted writing remains unsolved.

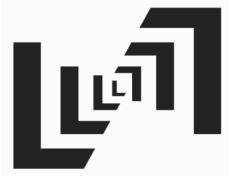

## 2 Methods and materials

### 2.1 Data

For our evaluation, we curated two data sets: First, an assessment set consisted of a meticulously garnered set of twenty-two scientific publications in the form of eleven matched samples. Specifically, we manually identified and extracted eleven publications where ChatGPT was either listed as a co-author or appropriately acknowledged in the text. As these publications self-evidently belong to the "LLM-Assisted" category, for each of these publications, a counterpart publication that was authored by the leading human author (i.e. first author) during the 2021- 2022 period was matched, resulting in eleven paired samples. Note that the publications chosen from the 2021-2022 period are assumed to be free of any LLM influences given that this period predates LLM proliferation (Lund & Wang, 2023). The full list of publications considered is provided in Appendix 1. Second, a false-positive set was assembled, comprising of a varied compilation of 1,094 publications published in or before 2022 (i.e. devoid of any LLM influences). The curation process, which is detailed in Appendix 2, follows a similar technique to thatpresented in recent literature (Alexi et al., 2024; Zargari et al., 2023). The resulting false-positive set consists of full-text manuscripts fromestablished authors across diverse academic institutions, disciplines, and ranks.

## 2.2 LLM-generated text detectors

We consider four state-of-the-art open-access LLM-generated text detectors: DetectLLM (Su et al., 2023), Zippy, LLMDet (Wu et al., 2023), and ConDA (Bhattacharjee et al., 2023). These four detectors exemplify the two predominant approaches in detecting LLM-generated text: zero-shot, represented by the first two detectors, meaning they do not necessitate additional input during inference other than the provided text of interest; and few-shot, represented by the latter two detectors, re- quiring a small number of reference samples for their inference. We utilized the implementations of these detectorsas originally published by their authors for our analysis. It is worth noting that two of the detectors offer a "soft classification", meaning they provide a continuous measure that requires conversion into a "hard classification", i.e. a binary label of "LLM-assisted" or not. We tune these decision threshold parameters using a simple grid search approach with the assessment set.

## 2.3 Alternative detector

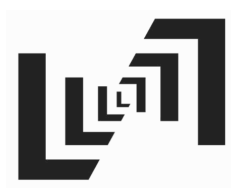

We further evaluate a simple writing style-based approach for detecting LLM-assisted writing. The approach is based on the premise that a sudden change in one's writing style around the time of LLM proliferation could potentially indicate LLM-assisted writing, especially if the change aligns with LLM writing style. Our detector, which we term the LLM-Assisted Writing (LAW) detector for simplicity, works as illustrated in Figure 1: First, for training, we adopt the writing style modeling technique provided by (Lazebnik & Rosenfeld, 2023), and for a given author $a$, we use themost recent publications made in or before 2022 (i.e. free of LLM influences) for modeling the author's writing style dynamics. Specifically, since one's writing style may vary over time regardless of LLM influences (Lazebnik & Rosenfeld, 2023), we measure the average change in the presented writing style from one publication to the next, and the standard deviation of this change, for the most recent six LLM-free publications, denoted $Avg(a)$ and $STD(a)$, respectively. Then, at the inference phase, for a given publication made in 2023 by $a$, we use a naive anomaly detection approach and consider the publication anomalous if its writing style significantly differs from $a$'s earlierpublications by at least $Avg(a) + STD(a)$. Note that slight variations to the above definitions, such as relying on a different number of prior publications (i.e. between 2 and 10) for computing $Avg(a)$ and $STD(a)$ and/or using a $Avg(a) + k \cdot STD(a)$ (with $k$ instead of $k=1$) bring about highly similar outcomes using our data and thus are not considered separately. For an identified anomaly, we compute the difference between its writing style vector and the average writing style vector computed for earlier works, resulting ina so-called "delta vector". Intuitively, this vector represents the unique characteristics of the given publication compared to earlier publications. To attribute these changes to LLM assistance we follow (Semrl et al., 2023), and provide an LLM of interest with the title and abstract of the publication, asking it to generate an academic manuscript, using the following query: "You are a scholar working on a new academic manuscript. The title of the manuscript is: <title-goes-here>. The abstract of the manuscript is: <abstract-goes-here>. Please write the entire manuscript." Once the LLM-written manuscript was obtained, we computed the cosine similarity between the delta vector and the writing style vector of the LLM-written text. Finally, if the similarly is higher than a given decision threshold parameter θ, the anomaly is classified as LLM-assisted writing. We tune this parameter using a grid search, as before.

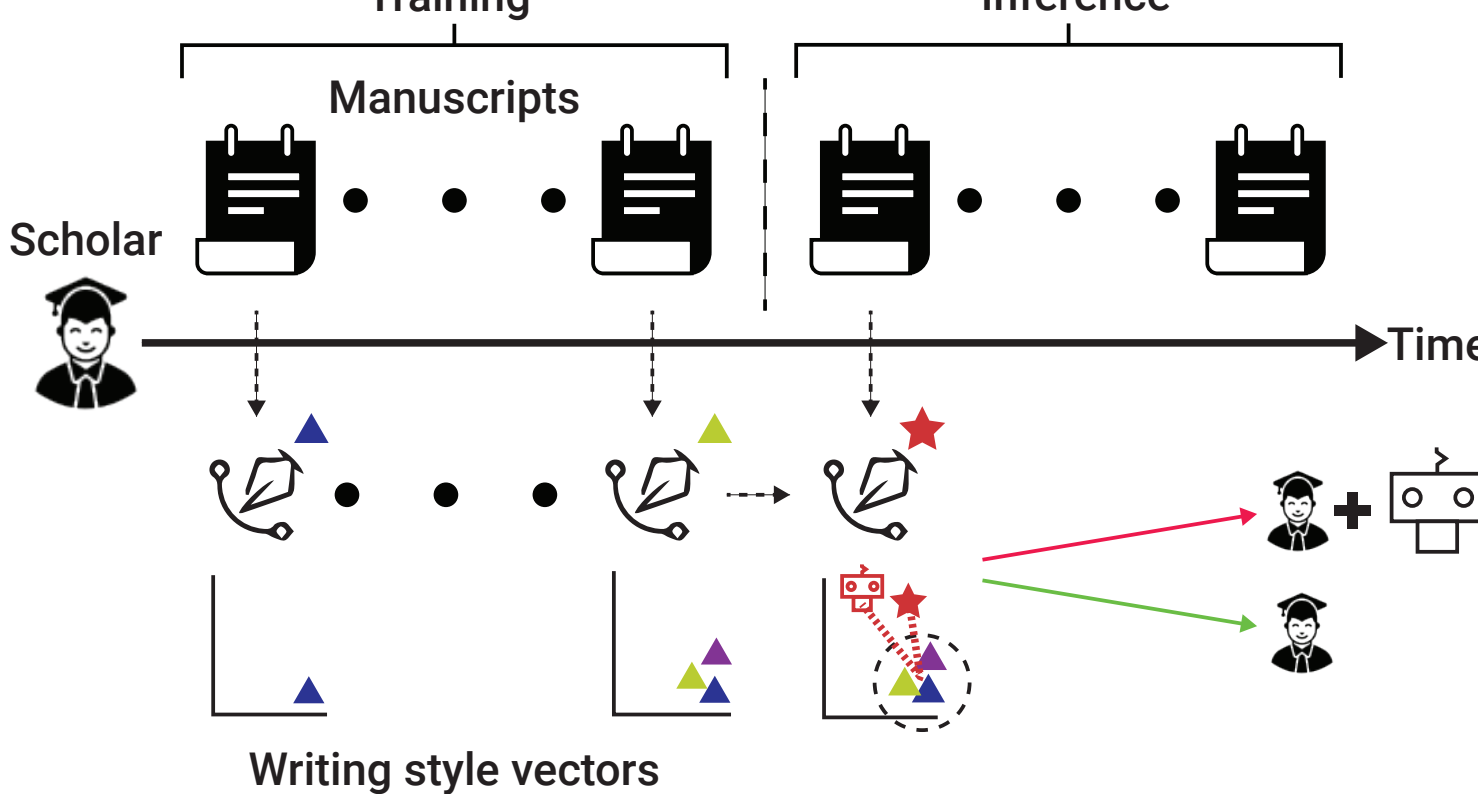

Figure 1. A schematic view of the LLM-Assisted Writing (LAW) detector. The detection process consists of twophases: First, during training, manuscripts are converted into vectors representing the author's writing style using the technique provided in (Lazebnik & Rosenfeld, 2023). The average change and standard deviation of the presented writing style are measured to capture the dynamics in one's writing style. Then, during inference, for each manuscript, we examine whether the change in its author's writing style is substantial enough to be considered an anomaly and whether this anomaly is aligned with the style of an LLM-generated manuscriptof the same title and abstract. If both conditions are met, the manuscript is deemed as an LLM-assisted manuscript.

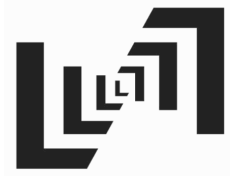

## 3 Results

Table 1 presents the results of each model in terms of accuracy, $F_1$ score, recall, precision, and false positive rate. Starting with the assessment data, the LAW detector favorably compares to the LLM-generated text detectors by providing a marginal improvement between 0.09 and 0.181 in accuracy, between 0.1 and 0.414 in terms of $F_1$ score, between 0.073 and 0.366 in terms of recall and between 0.125 and 0.45 in terms of precision. Similarly, for the false-positive set, the LAW detector favorably compares to the baseline detectors by providing a marginal improvement between 5.7% and 14.1% in the false positive rate.

Statistically, the five detectors do not differ significantly on the assessment set given its very limited size (11 paired samples). Nevertheless, the detectors statistically differ in their performance on the false positive set $\chi^2 = 133$, $p < 0.001$ with the LAW detector statistically outperforming three out of the four detectors at $p < 0.001$ following a Bonferroni post-hoc correction. The complete pair-wise comparison between the detectors, along with their agreement levels, are provided in Appendix 3.

Table 1. The performance of the examined detectors (columns) on the assessment set (first row) and the false- positive set (second row). The performance is presented as the accuracy with the $F_1$-score in brackets (for the assessment set) and as the false positive rate (for the false-positive set).

| Model | LLMDet | DetectLLM | ZipPy | ConDA | LAW |
|---|---|---|---|---|---|
| **Accuracy** | 0.546 | 0.591 | 0.637 | 0.637 | 0.727 |
| **F1-score** | 0.286 | 0.471 | 0.600 | 0.600 | 0.700 |
| **Recall** | 0.334 | 0.534 | 0.627 | 0.627 | 0.700 |
| **Precision** | 0.250 | 0.421 | 0.575 | 0.575 | 0.700 |
| **False Positive** | 17.2% | 13.8% | 9.7% | 8.8% | 3.1% |

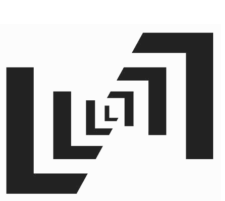

## 4 Discussion

The observed results suggest that existing state-of-the-art LLM-generated text detectors are suboptimal, at the very least, for the task of detecting LLM-assisted writing in scientific communication. This subpar performance is manifested as low accuracy, F1-scores, and high false positive rate, particularly when contrasted with the simplewriting style change detector we implemented in this study. We contend that these results should prompt a call for the development of specialized detectors, exclusively dedicated to LLM-assisted writing detection, aiming for more robust performance in the near future. We are of the opinion that such development is warranted and couldplay a pivotal role in fostering more authentic recognition of LLM-assisted writing and, consequently, it has the potential to enhance transparency, honesty, and integrity in scientific communication.

Our study is not without limitations. First, our ad-hoc writing style-based detector is consciously developed to detect an unexpected change in writing style during the time of LLM proliferation. As such, authors with a limited number of publications made prior to that period could not be considered by the detector. Moreover, mild changes in writing style from one publication to the next would likely not be detected, allowing for more substantial undetected LLM-assisted writing over time. In general, applying our detector for future publications would entail a major challenge of determining how and if to use current publications, be they classified by the detector as LLM-assisted or not, for future inference. Regarding our evaluation, it relies on two data sets with the first being relatively small. Unfortunately, at least in the realm of scientific communication, gathering more unsolicited instances of LLM-assisted writing is highly challenging since, as currently believed, most authors whopractice LLM-assisted writing avoid explicitly reporting it for a variety of reasons (Dergaa et al., 2023; Yuan et al., 2022).

## Funding information

This research did not receive any specific grant from funding agencies in the public, commercial, or not-for-profitsectors.

## Data availability

The data that has been used in this study is available by a formal request from the authors.

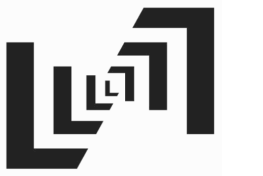

## Author contributions

Teddy Lazebnik (lazebnik.teddy@gmail.com): Conceptualization, Methodology, Software, Formal analysis, Investigation, Visualization, Project administration, Writing - Review & Editing.

Ariel Rosenfeld (ariel.rosenfeld@biu.ac.il): Conceptualization, Methodology, Investigation, Validation, Writing - Original Draft, Writing - Review & Editing.

## Statement of using AIGC

The authors used ChatGPT-4 during the writing of the manuscript to improve the wording of the text.

## Appendix

## 1 Assessment set: List of publications

Table A1. List of manuscripts included in the assessment set.

| LLM-assisted Writing | Counterpart |
|---|---|
| Osterrieder, J., GPTChat, A Primer on Deep Reinforcement Learning for Finance, SSRN (2023) | Finance, F., Osterrieder, J., Generative Adversarial Networks in finance: an overview, arXiv (2021) |
| Biswas, S., Will ChatGPT take my Job? Replies and Advice by ChatGPT, SSRN (2023) | Biswas, S., Role of Sonography in Ocular Trauma: A Study, ARC Journal of Surgery (2021) |
| Askr, H., Darwish, A., Hassanien, A.E., ChatGPT, The Future of Metaverse in the Virtual Era and Physical World: Analysis and Applications. Studies in Big Data (2023) | Gad, I., Hassanien, A. E., A wind turbine fault identification using machine learning approach based on pigeon inspired optimizer, Tenth International Conference on Intelligent Computing and Information Systems (2021) |
| King, M. R., chatGPT, A Conversation on Artificial Intelligence, Chatbots, and Plagiarism in Higher Education, Cellular and Molecular Bioengineering (2023) | King, M. R., CMBE Moves to the Structured Abstract Format: A Note from the Editor, Cellular and Molecular Bioengineering (2017) |
| Kung et al., Performance of ChatGPT on USMLE: Potential for AI-Assisted Medical Education Using Large Language Models, medRxiv (2022) | Kung, H. K., Host physician perspectives to improve predeparture training for global health electives, medical education (2017) |
| O'Connor S., Open artificial intelligence platforms in nursing education: Tools for academic progress or abuse?, Nurse Education in Practice (2022) | O'Connor S., Exoskeletons in Nursing and Healthcare: A Bionic Future, Clinical nursing research (2021) |
| Rossoni, L., A inteligencia artificial e eu: escrevendo o ˆ editorial juntamente com o ChatGPT, Revista Eletronica ˆ de Ciencia Administrativa (2022) | Rossoni, L., Editorial: A RECADM no Redalyc e o Dilema das Bases e Indexadores, Revista Eletronica de ˆ Ciencia Administrativa (2021) |
| chatGPT, Zhavoronkov, A., Rapamycin in the context of Pascal's Wager: generative pre-trained transformer perspective, Oncoscience (2022) | Zhavoronkov, A., The inherent challenges of classifying senescence, Science (2020) |
| Biswas, S., ChatGPT and the Future of Medical Writing, Radiology (2023) | Biswas, S., Biswas, S., A Study on penile doppler, MedCrave Online Journal of Surgery (2017) |
| Lazebnik, T., ChatGPT, The Impact of Fruit and Vegetable Consumption and Physical Activity on Diabetes Risk among Adults, arXiv (2022) | Lazebnik, T., Bunimovich-Mendrazitsky, S., The Signature Features of COVID-19 Pandemic in a Hybrid Mathematical Model—Implications for Optimal Work–School Lockdown Policy, Advanced Theory and Simulations (2021) |
| BaHammam, A. S., Trabelsi, K., Pandi-Perumal, S. R., Jahrami, H., Adapting to the Impact of AI in Scientific Writing: Balancing Benefits and Drawbacks while Developing Policies and Regulations, Journal of Nature and Science of Medicine (2023) | Akhtar, N., Ravi Gupta, S.R. Pandi-Perumal, Ahmed S. BaHammam: Clinical Atlas of Polysomnography: A Book Review, Sleep and Vigilance (2021) |

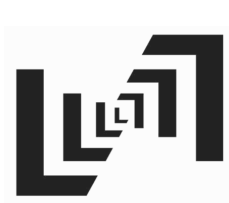

## 2 False-positive set: Curation process

We utilized 359 Google Scholar (GS) profiles, originally retrieved and analyzed by Zargari et al. (2023) for different research purposes. These profiles were sampled by the original authors from the top 200 U.S.-based institutions (based on the Shanghai Academic Ranking of National Universities ranking of 2022), covering five disciplines (Life Sciences, Exact Sciences, Law, Humanities, and Social Sciences) and five academic ranks (Adjunct Professor, Assistant Professor, Associate Professor, Full Professor, and Professor Emeritus). For the curation of the false-positive set, these 359 scholars served as "seed" points for a breadth-first search retrieval process through which we extracted these scholars' GS profiles, their co-authors' GS profiles, and their co-authors' co-authors' GS profiles.Through this process, a set of roughly 120 thousand GS profiles were retrieved. For evaluation purposes, we focus on profiles for which we have full access[1] to at least six publications made in or before 2022 (for training) andat least one additional publication made in or before 2022 (for inference). Overall, 1,094 profiles met the above requirements. For each of these profiles, the most recent publication made in or before 2022 was included in the false-positive set.

## 3 Further statistical analysis

### Pair-wise comparisons

Table A2. Pairwise comparison between the five detectors. The results are shown as *p value* with the statistics inbrackets. Each cell contains the results for the assessment set on the left, and the results for the false positive set on the right.

| | LLMDet | DetectLLM | ZipPy | ConDA |
|---|---|---|---|---|
| **DetectLLM** | 0.66(0.19)/< 0.01(10.45) | | | |
| **ZipPy** | 0.38(0.78)/< 0.01(69.63) | 0.66(0.20)/< 0.01(20.96) | | |
| **ConDA** | 0.06(3.67)/< 0.01(95.71) | 0.66(0.20)/< 0.01(34.21) | 1.0(0.0)/0.28(1.13) | |
| **LAW** | 0.01(0.03)/< 0.01(729.19) | 0.15(2.06)/< 0.01(34.21) | 0.34(0.92)/< 0.01(161.74) | 0.34(0.92)/< 0.01(120.46) |

### Agreement levels

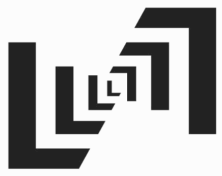

The Fleiss' κ calculated between the five detectors in question is 0.78. Table A3 reports the pairwise Cohan's κ.

[1] Through our institutions' publisher agreements or open access.

Table A3. Pairwise Cohan's κs calculated for the five detectors. Each cell contains the results for the assessment seton the left, and the results for the false positive set on the right.

| | DetectLLM | ZipPy | ConDA | LAW |
|---|---|---|---|---|
| **LLMDet** | 0.86 / 0.82 | 0.68 / 0.74 | 0.67 / 0.72 | 0.63 / 0.69 |
| **DetectLLM** | | 0.72 / 0.76 | 0.67 / 0.75 | 0.59 / 0.62 |
| **ZipPy** | | | 0.86 / 0.96 | 0.77 / 0.88 |
| **ConDA** | | | | 0.81 / 0.90 |

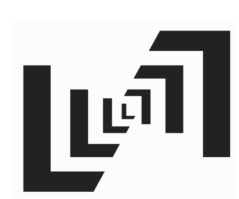

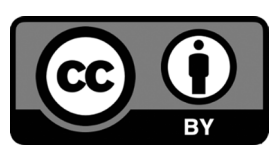

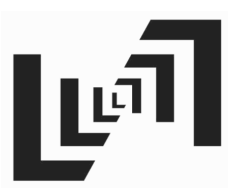